\documentclass[Physsubmission, Phys]{SciPost}

\hypersetup{
    colorlinks,
    linkcolor={red!50!black},
    citecolor={blue!50!black},
    urlcolor={blue!80!black}
}

\usepackage[bitstream-charter]{mathdesign}
\DeclareSymbolFont{usualmathcal}{OMS}{cmsy}{m}{n}
\DeclareSymbolFontAlphabet{\mathcal}{usualmathcal}

\def\BA{\begin{eqnarray}}
\def\BE{\begin{equation}}
\def\BF{\begin{figure}[htb]}
\def\BT{\begin{table}[htb]}
\def\EA{\end{eqnarray}}
\def\EE{\end{equation}}
\def\EF{\end{figure}}
\def\ET{\end{table}}

\def\fm{\,\mbox{fm}}

\def\mb{\,\mbox{mb}}

\def\sqq{\sigma_{Q\bar Q}}

\def\lsim{\mathrel{\rlap{\lower4pt\hbox{\hskip1pt$\sim$}}
     \raise1pt\hbox{$<$}}}         
 \def\gsim{\mathrel{\rlap{\lower4pt\hbox{\hskip1pt$\sim$}}
     \raise1pt\hbox{$>$}}}         

\begin{document}

\begin{center}{\Large \textbf{
Momentum transfer dependence of heavy quarkonium electroproduction\\
}}\end{center}

\begin{center}
M. Krelina\textsuperscript{1,2,*} and
J. Nemchik\textsuperscript{1,3} 
\end{center}

\begin{center}
{\bf 1} FNSPE, Czech Technical University in Prague, Brehova 7, 11519 Prague, Czech Republic
\\
{\bf 2} Physikalisches Institut, University of Heidelberg,
Im Neuenheimer Feld 226, 69120 Heidelberg, Germany 
\\
{\bf 3} Institute of Experimental Physics SAS, Watsonova 47, 04001 Kosice, Slovakia
\\
* michal.krelina@fjfi.cvut.cz
\end{center}

\begin{center}
\today
\end{center}


\definecolor{palegray}{gray}{0.95}
\begin{center}
\colorbox{palegray}{
  \begin{tabular}{rr}
  \begin{minipage}{0.1\textwidth}
    \includegraphics[width=22mm]{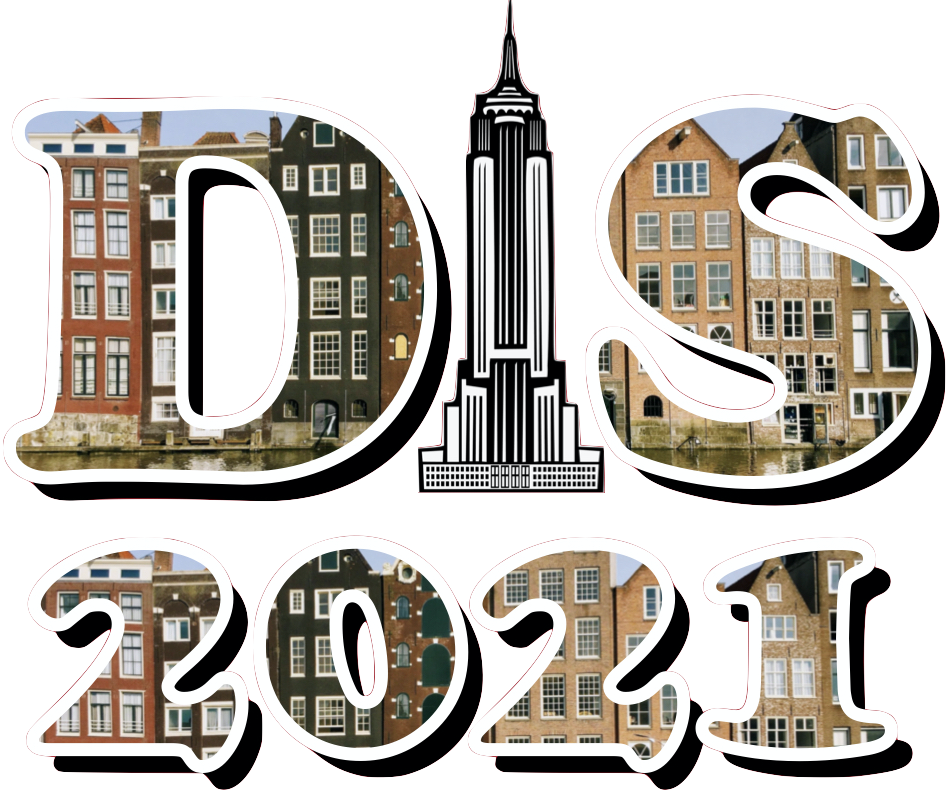}
  \end{minipage}
  &
  \begin{minipage}{0.75\textwidth}
    \begin{center}
    {\it Proceedings for the XXVIII International Workshop\\ on Deep-Inelastic Scattering and
Related Subjects,}\\
    {\it Stony Brook University, New York, USA, 12-16 April 2021} \\
    \doi{10.21468/SciPostPhysProc.?}\\
    \end{center}
  \end{minipage}
\end{tabular}
}
\end{center}

\section*{Abstract}
{\bf
We investigate the momentum transfer dependence of differential cross sections $d\sigma/dt$ in diffractive electroproduction of heavy quarkonia on proton targets.
Model predictions for $d\sigma/dt$ within the light-front QCD dipole formalism are based on a realistic model for a proper correlation between the impact parameter $\vec b$  of a collision and color dipole orientation $\vec r$.
We demonstrate a significance of $\vec b-\vec r$ correlation by comparing with a standard simplification $\vec{b}\parallel\vec{r}$, frequently used in the literature.
}

%
\vspace{10pt}

\vspace*{-0.3cm}
%
%
%
\section{Introduction}
\label{sec:intro}
%
%
%

Photo- and electroproduction of heavy quarkonia represents a unique tool allowing to study diffraction mechanism, saturation phenomena, gluon distribution functions, etc. 
However, for a proper analysis of a given effect associated with the corresponding electroproduction process, it is essential to know various theoretical uncertainties, such as the $Q-\bar Q$ interaction potential, which generates quarkonium wave functions \cite{Cepila:2019skb}, the $Q \bar Q \to V$ vertex structure in connection with an open question about a contribution of the $D$-wave component in quarkonium wave functions \cite{Krelina:2019egg}, or the shape of dipole cross section $\sqq$, which represents one of the main ingredients within the color dipole formalism \cite{Cepila:2019skb,Kopeliovich:2020has}. 

Theoretical and experimental investigations of the transverse momentum transfer $\vec q$ dependence of differential cross sections $d\sigma/dq^2$ provide the opportunity for a more detailed study of the QCD dynamics accompanying the diffractive quarkonium production. Knowledge of $\vec q$-orientation leads to an identification of the reaction plane since $\vec q$ is related to the impact parameter $\vec b$ of a collision via Fourier transform. Performing model predictions, this generates a task to include a correlation between dipole orientation $\vec r$ and the vector $\vec b$ properly. Such $\vec b-\vec r$ correlation is not implemented adequately in most models for $b$-dependent dipole cross sections. In the present paper we rely on the model from \cite{Kopeliovich:2021dgx} and analyze how the accurate $\vec b-\vec r$ correlation modify the results of $d\sigma/dq^2$ by comparing with predictions based on a simplified assumption $\vec{b}\parallel\vec{r}$. 

The next Sec.~\ref{sec:dipFrm} contains a short introduction to the color dipole formalism. 
The explicit form of the partial $Q\bar Q$-proton amplitude with a proper $\vec b-\vec r$ correlation is presented in Sec. ~\ref{sec:dipAmpl}. The impact of such $\vec b-\vec r$ correlation on magnitudes and shape of $d\sigma/dq^2$ is analyzed in Sec.~\ref{sec:res}. Our main results are summarized in Sec.~\ref{sec:concls}.

\vspace*{-0.2cm}
%
%
%
\section{Color Dipole Framework}
\label{sec:dipFrm}
%
%
%

Within the light-front (LF) color dipole formalism, the amplitude for electroproduction of heavy vector mesons with the transverse momentum transfer $\vec q$ can be expressed in the factorized form,
%
\begin{equation}
\mathcal{A}^{\gamma^\ast p\to V p}(x,Q^2,\vec q)
=
\bigl\langle V |\tilde{\mathcal{A}} |\gamma^*\bigr\rangle
=
\int d^2r\int_0^1 d\alpha\,
\Psi_{V}^{*}(\vec r,\alpha)\,
\mathcal{A}_{Q\bar Q}(\vec r, x, \alpha,\vec q)\,
\Psi_{\gamma^\ast}(\vec r,\alpha,Q^2)\,,
\label{eq:amp-p0}
\end{equation}
%
where $\mathcal{A}_{Q\bar Q}(\vec r, x, \alpha,\vec q)$ is the amplitude for elastic scattering of the color dipole
on the nucleon target,
$\Psi_V(r,\alpha)$ is the LF wave function for heavy quarkonium
and $\Psi_{\gamma^\ast}(r,\alpha,Q^2)$ is the LF distribution of the $Q\bar Q$ Fock component of the real ($Q^2 = 0$) or virtual ($Q^2 > 0$) photon,
where $Q^2$ is the photon virtuality and the $Q\bar Q$ fluctuation (dipole) has the transverse size $\vec{r}$.
The variable $\alpha$ is the fractional LF momentum carried by a heavy quark or antiquark from a $Q\bar Q$ Fock component of the photon and $x =  (m_V^2 + Q^2 -t)/(W^2 + Q^2 - m_N^2)\approx (m_V^2+Q^2-t)/s$, where $m_V$ and $m_N$ is the quarkonium and nucleon mass, respectively, $W$ is c.m. energy of the photon-nucleon system and $t = -q^2$.

Most of phenomenological studies of the partial dipole amplitude $\mathcal{A}_{Q\bar Q}$ are performed in the impact parameter representation, where $b$-dependent amplitude $\mathcal{A}_{Q\bar Q}(\vec r, x, \alpha,\vec b)$ is related to $\mathcal{A}_{Q\bar Q}(\vec r, x, \alpha,\vec q)$ in Eq.~(\ref{eq:amp-p0}) via Fourier transform,
%
\begin{equation}
\mathcal{A}_{Q\bar Q}(\vec r, x, \alpha,\vec q)
=
\int d^2b \, e^{-i \vec b \cdot \vec q}
\mathcal{A}_{Q\bar Q}(\vec r, x, \alpha,\vec b)
\end{equation}
%
with the correct reproduction of the dipole cross section at $\vec q=0$
%
\begin{equation}
    \sqq(r,x) = \textrm{Im} \mathcal{A}_{Q\bar Q}(\vec r, x, \alpha,\vec q=0) = 2 \int d^2b \, 
    \textrm{Im}
    \mathcal{A}_{Q\bar Q}^N(\vec r, x, \alpha,\vec b)\,,
    \label{eq:b-integ}
\end{equation}
%
where the partial dipole amplitude $\mathcal{A}_{Q\bar Q}^N(\vec r, x, \alpha,\vec b)$ represents the interaction of the $Q \bar Q$ dipole acquiring orientation $\vec r$ with a nucleon target at the impact parameter $\vec b$.

The exclusive electroproduction differential cross section on a proton target reads, 
%
\begin{equation}
\frac{d\sigma^{\gamma^{\ast} p\to V p}(s,Q^2,t=-q^2)}{dt}
=
\frac{1}{16\,\pi}\,
\Bigl |
\mathcal{A}^{\gamma^{\ast} p\to V p}(s,Q^2, \vec q)
\Bigr |^2,
\label{eq:proton}
\end{equation}
%
where we adopt wave functions of quarkonia generated by the realistic Buchmuller-Tye (BT) $Q-\bar Q$ interaction potential and a simple "S-wave-only" $V\to Q\bar Q$ transition requiring to perform the Melosh spin rotation \cite{Cepila:2019skb,Kopeliovich:2021dgx}. The corresponding final formulas for differential cross sections can be found in \cite{Kopeliovich:2021dgx}.

\vspace*{-0.2cm}
%
%
%
\section{Partial Dipole Amplitude}
\label{sec:dipAmpl}
%
%
%

\begin{figure}[ht]
\centering
\includegraphics[width=0.85\textwidth]{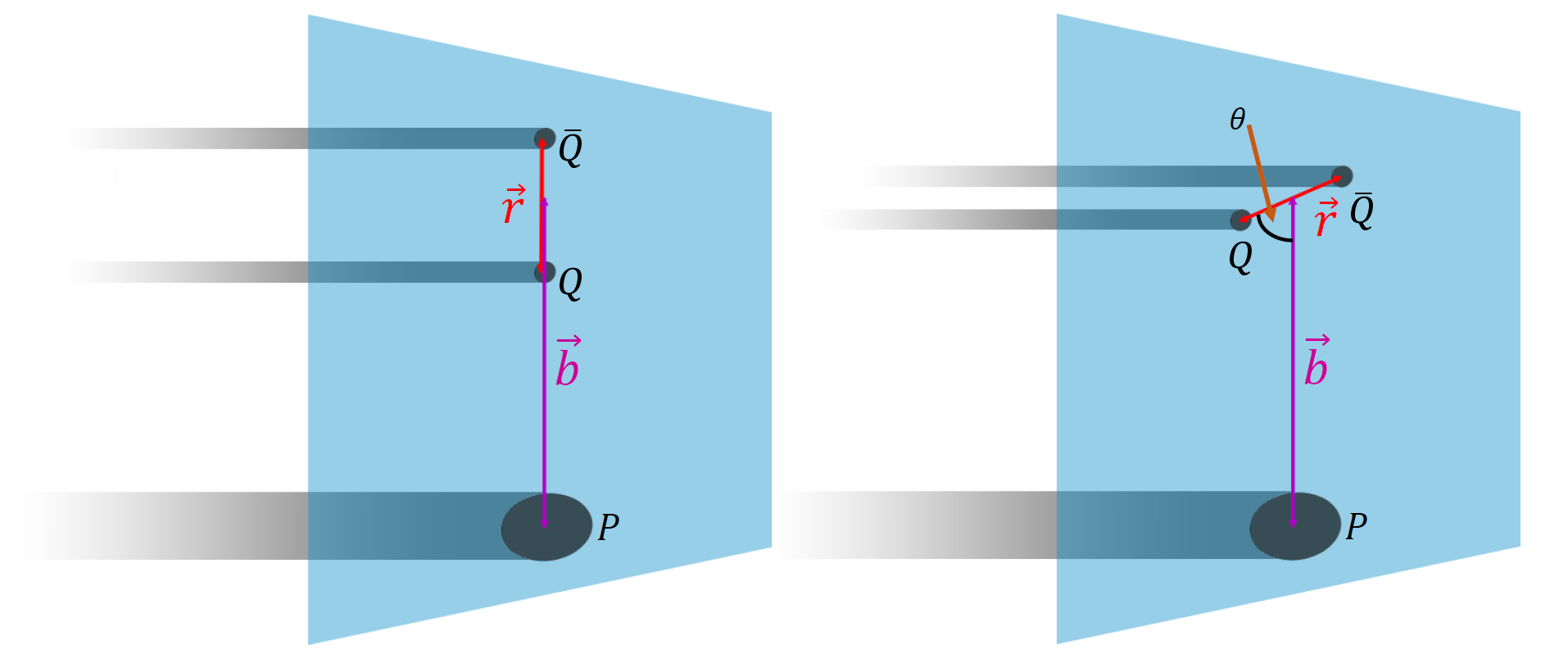}
\caption{A cartoon demonstrating a significance of dipole orientation. Whereas the left panel illustrates the standard approximation $\vec{r}$$\parallel$$\vec{b}$ (an angle between vectors $\vec{r}$ and $\vec{b}$ is fixed at $\theta = 0^o$), the right panel represents the general case with no restrictions for an angle $\theta$ requiring so a subsequent integration over $\theta$ in calculations.
}
\label{fig:planes}
\end{figure}

Let's be $\vec r$ the transverse separation of a colorless heavy quark $Q\bar Q$ photon fluctuation (dipole) and the vector $\vec b$ is the impact parameter of its centre of gravity. Then the corresponding
interaction of the $Q\bar Q$ dipole is possible due to the difference between impact parameters of $Q$ and $\bar Q$ relative to the scattering centre. Thus, independently of the magnitude of $\vec r$, the production of any $Q\bar Q$ component with the same impact parameter from the target related to $Q$ and $\bar Q$ is terminated (see Fig.~\ref{fig:planes}). This leads to vanishing and maximal dipole interaction if $\vec b\bot\vec r$ and $\vec b\parallel\vec r$, respectively. 

Such a correlation between the vectors $\vec b$ and $\vec r$ is incorporated in the partial elastic dipole amplitude $A_{Q\bar Q}^N(\vec r,x,\alpha,\vec b)$ introduced in \cite{Kopeliovich:2007fv}
within the standard model for the dipole cross section of a conventional saturated form,
$
\sqq(r,x) =
\sigma_0\,
\left(1 - \exp \left[ - r^2/R_0^2(x)\right] \right)
$.
The corresponding partial dipole apmlitude reads \cite{Kopeliovich:2007fv,Kopeliovich:2008nx,Kopeliovich:2021dgx},
%
\begin{eqnarray}
\mathrm{Im} \mathcal{A}^N_{Q\bar Q}(\vec r, x, \alpha,\vec b\,)
=
\frac{\sigma_0}{8\pi \mathcal{B}(x)}\,
\Biggl\{
\exp\left[-\,\frac{\bigl [\vec b+\vec
r(1-\alpha)\bigr ]^2}{2\mathcal{B}(x)}\right] 
+ 
\exp\left[-\,\frac{(\vec
b-\vec r\alpha)^2}{2\mathcal{B}(x)}\right]
\nonumber\\
- \,2\,\exp\Biggl[-\,\frac{r^2}{R_0^2(x)}
-\,\frac{\bigl [\,\vec b+(1/2-\alpha)\vec
r\,\bigr ]^2}{2\mathcal{B}(x)}\Biggr]
\Biggr\}\,,
\label{eq:dipa-gbw}
\end{eqnarray}
%
where $\mathcal{B}(x)=R_N^2+R_0^2(x)/8$ with $R_N^2$ 
related to the constant term in the standard Regge parametrization for the energy-dependent $t$-slope of the differential elastic cross section.
In our calculations we adopt  the GBW dipole model \cite{GolecBiernat:1998js}, where the above parameters read: $\sigma_0 = 23.03\,\mb$,
$R_0(x) = 0.4\,\fm\times(x/x_0)^{0.144}$ with $x_0 = 3.04\times 10^{-4}$.
The form of the dipole amplitude (\ref{eq:dipa-gbw}) correctly reproduces at $\vec q=0$ the dipole cross section according to Eq.~(\ref{eq:b-integ}).

\vspace*{-0.2cm}
%
%
%
\section{Results and discussions}
\label{sec:res}
%
%
%

\begin{figure}[ht]

\includegraphics[scale=0.72]{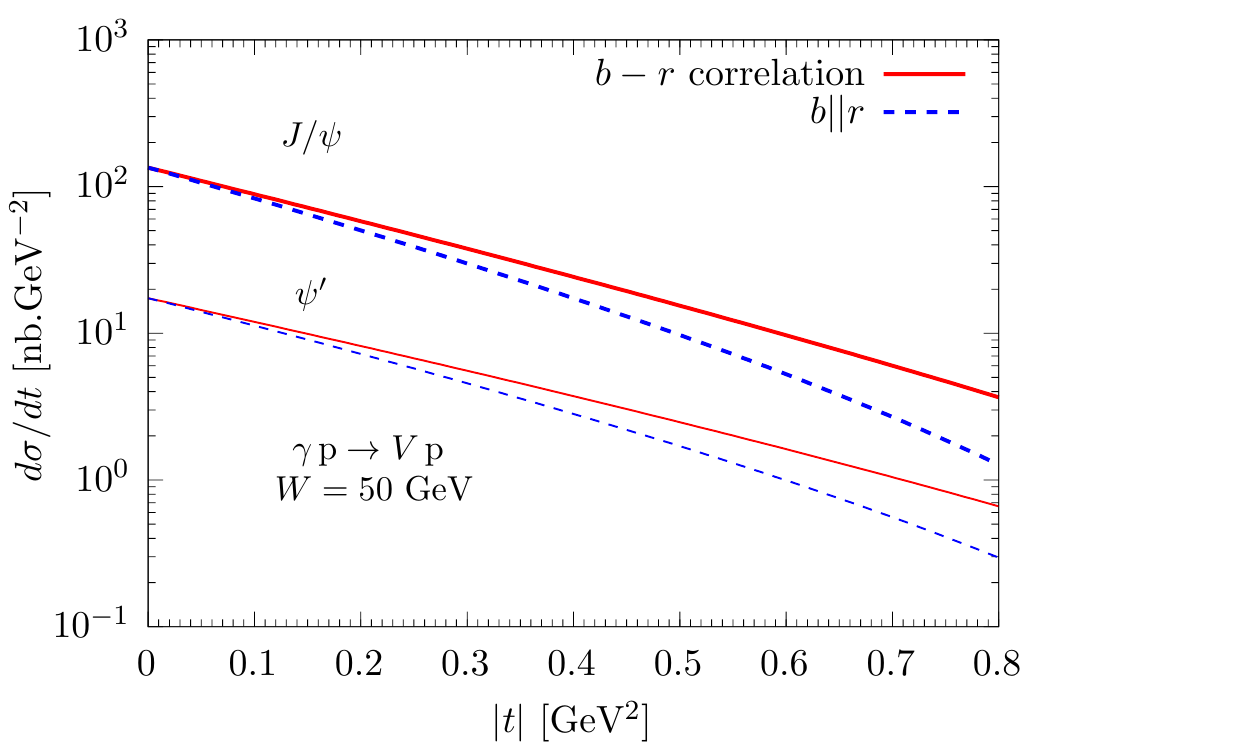}
\hspace*{-1.6cm}
\includegraphics[scale=0.72]{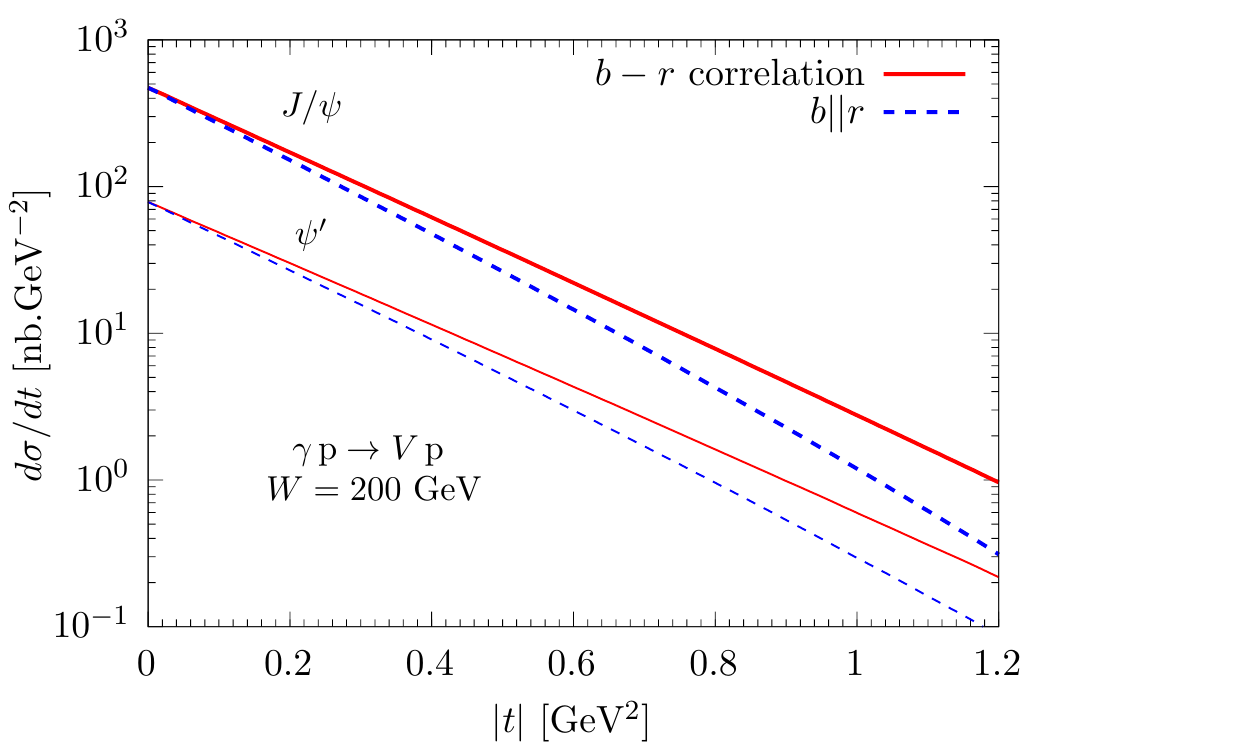}\\
\includegraphics[scale=0.72]{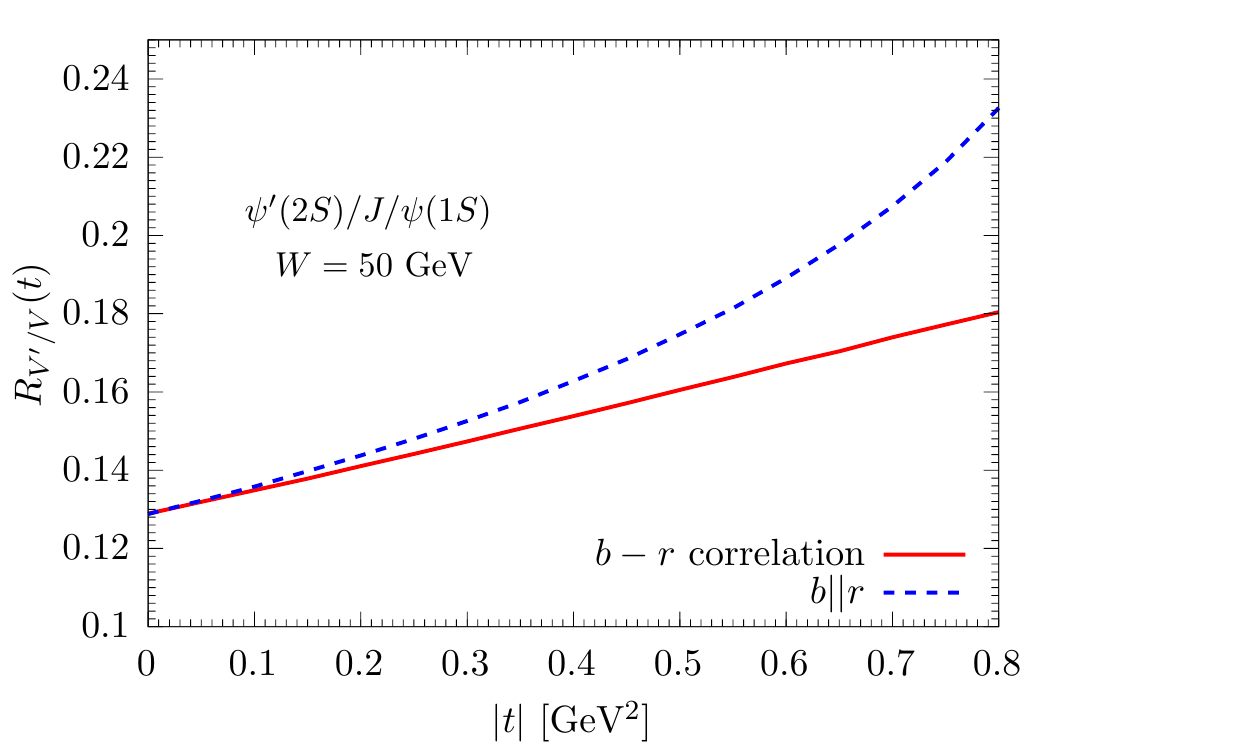}
\hspace*{-1.6cm}
\includegraphics[scale=0.72]{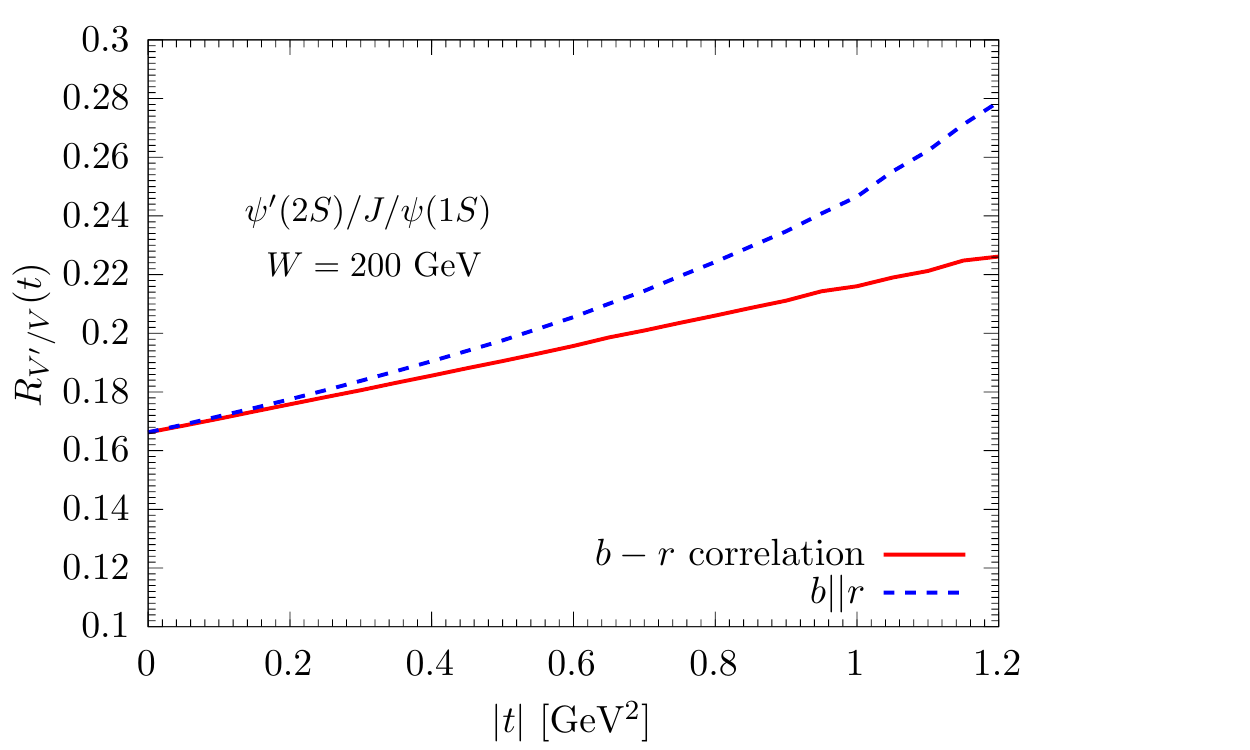}
\caption{
  \label{fig:Fig-tdep-corr}
  Manifestation of an importance of a proper $\vec b - \vec r$ correlation in the partial elastic dipole amplitude by performing calculations of differential cross sections $d\sigma^{\gamma p\to J/\psi (\psi') p}(t)/dt$ (top panels) and $\psi'(2S)$-to-$J/\psi(1S)$ ratio of differential cross sections (bottom panels)  at c.m. energies $W=50$ (left panels) and $200$~GeV (right panels).
  }
\end{figure}

To quantify a significance of the correlation between vectors $\vec b$ and $\vec r$, we compare our calculations with those based on the simplified assumption $\vec b\parallel\vec r$. This is shown in Fig.~\ref{fig:Fig-tdep-corr}, which demonstrates an importance of the dipole orientation in the partial amplitude $\mathcal{A}^N_{Q\bar Q}$, Eq.~(\ref{eq:dipa-gbw}), (solid lines) with respect to the $\vec b\parallel\vec r$ case (dashed lines) at c.m. energies $W=50$ (left panel) and $200$~GeV (right panel).
In order to study a net effect of $\vec b - \vec r$ correlation itself, we omit in calculations the corrections for the real part of the production amplitude and the skewness effect.

One can see from top panels of Fig.~\ref{fig:Fig-tdep-corr} that our model predictions for $d\sigma/dt$ including a realistic $\vec b - \vec r$ correlation in the partial dipole amplitude (\ref{eq:dipa-gbw})
differ significantly from results based on a simplified assumption when $\vec b\parallel\vec r$ (see differences between solid and dashed lines).
The corresponding $t$-slopes of $d\sigma/dt$ are rather different, what has an indispensable impact on all predictions where authors assume that the dipole amplitude is independent of the angle between vectors $\vec b$ and $\vec r$. 
The significance of dipole orientation rises towards smaller photon energies corresponding to the energy range of experiments at the LHC and future electron-ion colliders. This gives a possibility 
to eliminate various models for $b$-dependent dipole amplitude from the description of diffractive quarkonium electroproduction.

The node effect in production of the $\psi'(2S)$ state can be investigated through the $t$-dependent $\psi'(2S)$-to-$J/\psi(1S)$ ratio $R_{\psi'/J/\psi}(t)\equiv R(t)$ of differential cross sections as is demonstrated in bottom panels of Fig.~\ref{fig:Fig-tdep-corr}. One can see that it causes a rather steep rise of $R(t)$, which is gradually changed for a more flat $t$-behavior at higher photon energy due to a weaker node effect. Let's suppose that $\vec b-\vec r$ correlation in the dipole amplitude is not included properly. In that case, we predict a much stronger rise with $t$ of the ratio $R(t)$ especially at the smaller photon energy (see differences between dashed and solid lines in bottom panels of Fig.~\ref{fig:Fig-tdep-corr}). This represents another way of ruling out various $b$-dependent models describing the partial dipole amplitude.

\vspace*{-0.2cm}
%
%
%
\section{Conclusions}
\label{sec:concls}
%
%
%

We studied the impact of a proper $\vec b-\vec r$ correlation in the partial dipole amplitude on magnitudes of differential cross sections $d\sigma/dt$ for diffractive electroproduction of heavy quarkonia on proton targets.

We demonstrated that, in comparison with a correct $\vec b-\vec r$ correlation, usual approximation $\vec b\parallel\vec r$
leads to a larger $t$-slope of $d\sigma/dt$ and causes much steeper rise with $t$ of $\psi'(2S)$-to-$J/\psi(1S)$ ratio of differential cross sections. 

The significance of dipole orientation becomes stronger towards smaller photon energies and can be tested by experiments at the LHC and future electron-ion colliders.

\vspace*{-0.3cm}
%
%
%
\section*{Acknowledgements}
%
%
%

The work of J.N. was partially supported by Grant
No. LTT18002 of the Ministry of Education, Youth and
Sports of the Czech Republic,
by the project of the
European Regional Development Fund No. CZ.02.1.01/0.0/0.0/16\_019/0000778
and by the Slovak Funding Agency, Grant No. 2/0007/18.
The work of M.K. was supported by the International Mobility of Researchers - MSCA IF IV at CTU in Prague CZ.02.2.69/0.0/0.0/20\_079/0017983, Czech Republic.



\vspace*{-0.2cm}
%
%
%



\nolinenumbers

\end{document}